\documentclass[a4paper]{article}
\usepackage{graphicx}
\newcommand{\be}{\begin{equation}}
\newcommand{\ee}{\end{equation}}
\newcommand{\bd}{\begin{displaymath}}
\newcommand{\ed}{\end{displaymath}}
\newcommand{\bea}{\begin{eqnarray}}
\newcommand{\eea}{\end{eqnarray}}
\newcommand{\bi}{\begin{description}}
\newcommand{\ei}{\end{description}}
\newcommand{\bq}{\begin{quote}}
\newcommand{\eq}{\end{quote}}

\def\fo{\footnote}

\def\r{\rho}
\def\a{\alpha}

\def\g{\gamma}

\def\e{\epsilon}

\def\t{\tau}

\def\Om{\Omega}

\def\S{\Sigma}

\def\ph{\phi}

\begin{document}
\bibliographystyle{unsrt}
\author{Alexander~Unzicker\\
        Pestalozzi-Gymnasium  M\"unchen, Germany \\[0.6ex]
{\small{\bf e-mail:}  alexander.unzicker et lrz.uni-muenchen.de}}

\title{What Happened if Dirac, Sciama and Dicke had Talked to Each Other~?}
\title{How Cosmology Could Evolve if Dirac, Sciama and Dicke had Talked to Each Other}
\title{Should we Take Dirac Seriously~?}
\title{A Note on the  Contributions to Cosmology of \\
Dirac, Sciama and Dicke}
\title{What Happens to Cosmology by Combining Ideas of \\
Dirac, Sciama and Dicke}
\title{What Could Have Happened to Cosmology if \\
Dirac, Sciama and Dicke had Talked to Each Other}
\title{If Dirac, Sciama and Dicke had Talked to Each Other - A Trial to Encompass Their
Abandoned Contributions to Cosmology}
\title{How Cosmology Might Have Evolved Differently if \\ Dirac, Sciama and Dicke had Talked to Each Other}
\title{A Look at the Abandoned Contributions to Cosmology of\\ Dirac, Sciama and Dicke}
\maketitle

\begin{abstract}
The separate contributions to cosmology of the above  researchers are
revisited and a cosmology encompassing their basic ideas is
proposed. We study Dirac's article on the large number hypothesis
(1938), Sciama's proposal of
realizing Mach's principle (1953), and 
Dicke's considerations (1957) on a flat-space representation of general relativity
 with a variable speed of light (VSL). Dicke's 
tentative theory can be formulated
in a way which is compatible with Sciama's hypothesis on the
gravitational constant $G$. Additionally, such a
cosmological model is shown to satisfy Dirac's {\em second\/} `large number' hypothesis on
the total number of particles in the universe being proportional
to the square of the epoch. In the same context, Dirac's {\em
first\/} hypothesis on an epoch-dependent $G$
-contrary to his prediction- does not necessarily  produce a visible 
 time dependence of $G$. While Dicke's proposal
reproduces the classical tests of GR in first approximation,
the cosmological redshift is described by a shortening of measuring rods rather than
an expansion of space. Since  the temporal evolution of the horizon $R$ is governed by
$\dot R(t) =c(t)$, the flatness and horizon problems do not arise in the common form.

\end{abstract}

\section{Introduction}

Cosmology as a modern observational science started in the 1930s,
when Hubble's observations set an end to the `great debate'
whether andromeda is a nebula inside the milky way or an
independent  galaxy. Shortly after, with the first mass and
distance estimates of the whole universe, fundamental questions
regarding the interrelation of the universe with
 elementary particles were raised by Eddington and Dirac
\cite{Dir:37}. However, the main interest in Hubble's
redshift-distance law was due to the fact that it paved the way
for Friedmann-Lemaitre (FL) cosmology which naturally accompanied
the spectacular success of general relativity developed just 15
years before. As if cosmology liked great debates, the discussion
was then dominated by the rivalry between steady-state and
big-bang  models, which was decided in favor of the latter by the
discovery of the cosmic microwave
background (CMB). Notwithstanding such remarkable advance, there were 
reflective voices like Dirac's in 1968 \cite{Dir:68}:

\bq
`One field of work in which there has been too much speculation is cosmology.
There are very few hard facts to go on but theoretical workers have been busy
constructing various models for the universe based on any assumptions that
they fancy. These models are probably all wrong. It is usually assumed that
the laws of nature have always been the same as they are now. There is no
justification for this. The laws may be changing, and in particular, quantities
which are considered to be constants of nature may be varying with cosmological
 time. Such variations would completely upset the model makers.'
\eq Fortunately, cosmology in the meantime has many observational
facts which allow to do much more quantitative science than in the
1930's and even the 1960's. Dirac's prediction on a change of the
gravitational constant $G$ he had expressed in a manner slightly
too self-assured in his 1938 paper (`A new basis for cosmology'
\cite{Dir:38a}) has not been confirmed, GR
 has undergone an impressive series of confirmations, and FL- cosmology has
won all battles so far. This success and the lack of alternatives however bears the danger
of interpreting new data assuming a model we still should not forget to test.
Though having won all
battles, this was due to introducing considerable elasticity in the originally rigid theory by
means of dark matter, dark energy, and further numbers we must call fitting parameters
poorly understood so far. A review of the observational evidence for standard cosmology
and its problems is given elsewhere \cite{Unz:07}.

The great debates and the enthusiasm about new data of the past
decades brushed aside the interest in old unresolved problems and
it is a somehow unfortunate development that Dirac's criticism has
not been considered any more by theoreticians. Doubts on our usual
assumptions of the time-independency  of physical laws have been
expressed by various researchers \cite{Bar:88}: 
\bq 
The question
if there is a unique absolute standard of time which globally is
defined by the inner geometry of the universe, is a big unresolved
problem of cosmology.'\fo{Retranslated from the German edition.}
\eq 
In particular, the idea of time as an invisible river that
runs without relation to the universe (`time is what happens when
nothing else does') may be just wrong \cite{Bar}. A redefinition
of time by means of parameters which govern the evolution of the
universe should have profound consequences, though observational
evidence may be minute. Relating our local physical laws which
base on apparently constant quantities to global properties of the
universe is the greatest challenge of cosmology.

It should be clear that dealing with any alternative approach to
cosmology requires much patience, and a reinterpretation of all
new data we are flooded with cannot be done immediately.
Contrarily, we shall prepare ourselves to stay for a little while
in the period in which the unresolved problems first arose and
some old seminal papers were trying to understand them. A closer
look to their content is still fascinating: Dirac's large number
hypothesis, which consists of two independent, struggling
coincidences is one of the most mysterious, unexplained phenomena
in cosmology. He considered them as `fundamental though as yet
unexplained truths', which remain valid, even though the Hubble
"constant" varies with the age of the universe' \cite{Wei:72}. No
theoretical approach besides Jordan's one \cite{Jor:59} has taken them
seriously so far. Sciama's efforts to link the value of $G$ to the
mass distribution of the universe are accompanied by profound
insights and for the first time realized Mach's principle
concretely. Though Mach's ideas have always been as fascinating
and convincing from a conceptual point of view, the missing
quantitative formulation had remained an unsatisfying aspect. One
of the most interesting proposals in this context, full of
speculative ideas, has been given by Dicke \cite{Dic:57}, though
this is much less known than the later developed
scalar-tensor-theory. We shall first have a closer but compressed
look at the mentioned papers, and then give a technical
description of the proposal that makes use of the basic ideas
expressed there. First consequences are discussed in section~4.

\section{Review of the seminal papers}

\subsection{Dirac's Large Number Hypothesis}

Strictly speaking, Dirac's article \cite{Dir:38a},
relating three large dimensionless numbers occurring in physics,
expresses three different coincidences we shall name Dirac 0, I, and II.
Eddington had already noticed the number
\be
\frac{F_e}{F_g} = \frac{e^2 }{4 \pi \e_0 G m_p m_e} \approx 10^{39} \label{force}
\ee
and wondered how such a huge number could come out from any reasonable mathematical
theory. Dirac then observed that the age of the universe is about the same
multiple of the time light needs to pass the proton radius, or equivalently
\be
\frac{R_u}{r_p} \approx 10^{40} :=\e,  \label{epoch}
\ee
thus escaping from the mathematical difficulty of producing $\e$ and
postulating epoch-dependent forces. It was the first time somebody tried to relate
properties of the atomic scale to those of the universe as a whole. Not enough here, 
he noted the total number of baryons\fo{This argument is not changed
by the variety of elementary particles discovered in the meantime, since it 
involves orders of magnitude only.} being
\be
\frac{M_u}{m_p} \approx N \approx 10^{78} \approx \e^2. \label{particles}
\ee
As a consequence, for the gravitational constant, the relation
\be
G \approx \frac{c^2 R_u}{M_u} \label{mach}
\ee
must hold, a coincidence that was previously noted by Eddington and much earlier
(though lacking data, not in an explicit way) suggested by Ernst Mach, who insisted that
the gravitational interaction must be related to the presence of all masses
in the universe \cite{Mach}. 
I shall call the coincidences (\ref{force}+\ref{epoch}) Dirac I and
 (\ref{epoch}+\ref{particles}) Dirac II, while (\ref{mach})
should be named Dirac 0, to emphasize that Dirac's considerations
go much further: (\ref{mach}) could be realized either with a
different radius of elementary particles (not satisfying Dirac I)
or with another number of baryons of different weight (and being
in conflict with Dirac II).\fo{We shall not go into detail
regarding the question how Dirac II can be related to $h \approx c
m_p r_p$ and to the so-called Eddington-Weinberg number. The
agreement of the Compton wavelength of the proton with its actual
radius (determined by Rutherford) is however not trivial, as the
comparison with the electron shows.} It seems that Dirac himself
was more convinced of hypothesis I than of II. According to
\cite{Jor:59}, he abandoned the latter after various critiques,
e.g. by \cite{Fie:56}. Indeed, while Dirac I had a great influence
on physics with a huge amount of experimental tests (see
\cite{Uza} for an overview of the $\dot G \neq 0$ search)
contesting the appreciation of the idea, Dirac II remained
completely out of any theoretical approach so far. While Dirac I
would be fairly compatible with standard FL cosmology, Dirac II is
in explicit conflict with it. To be  concrete, FL cosmology
assumes for the epoch $\e = 10^{25}$ (BBN, creation of light
elements) a horizon containing $10^{64}$ baryons, while in the
epoch $\e = 10^{50}$ still $10^{78}$ (or, considering the
accelerated expansion, even less \cite{Krs:07}) baryons should be
seen. Dirac instead argued that `Such a coincidence we may presume
is due to some deep connection in nature between cosmology and
atomic theory.' (\cite{Dir:38a}, p.~201).
Jordan \cite{Jor:59} in 1959 commented 
on the second hypothesis:
\bq
`As far as I can see, I am the only one who was ready to take seriously Dirac's
model of the universe which was immediately abandoned by its creator. I have
to confess that I consider Dirac's thought as one of the greatest insights
of our epoch, whose further investigation is one of the big tasks.'
\eq
According to the predominant opinion among cosmologists however 
$N \sim \e^2$ is just a coincidence invented by nature to fool today's physicists.

Dirac was aware that a cosmological theory of this type could require
a change of time scales. In the sections~3 and~5 of his paper \cite{Dir:38a}, 
he considered an idealized time $t$ representing the epoch and an
observable time `$\t$' which were quadratically related and
considered\fo{See \cite{Wei:72}, eq.~16.4.6, for a comment on
that.} an evolution of the horizon $R \sim t^{\frac{1}{3}}$. Since
time measurements necessarily involve frequencies of atomic
transitions and therefore the speed of light, it is strange that
he maintained the postulate $c=1$. This omission led him to the
inviting but somewhat premature claim that the gravitational
constant $G$ had to vary inversely with the epoch. The amount of
experimental research done due to that prediction \cite{Uza}
illuminates the great influence of Dirac's ideas on physicists.
The so far (negative) outcome of the $\dot G \neq 0$ search has prevented
theorists from taking that deep principles too seriously, without
however having challenged the prediction as such from a theoretical
point of view.

\subsection{Sciama's implementation of Mach's principle}
Contrarily to Dirac, Sciama \cite{Sci:53} focussed on the question how to
realize Mach' principle in a quantitative form, having noticed that in Newton's theory
the value of $G$ is an arbitrary element (p.~39 below). From considerations we
skip here he derived a dependence of the gravitational constant\fo{eq.~(1) and~(5a)
with a change of notation.}
\be
G=\frac{c^2}{\sum_i \frac{m_i}{r_i}}, \label{sci}
\ee
whereby the  sum is taken over all particles and $r_i$ denoting
the distance to particle $i$. This is much more concrete and
quantitative than Mach's ideas or the speculations
of Eddington and Dirac. It provides further a reasonable dependency on distance
and alleviates the somewhat mysterious property that $G$ should `feel' the
whole universe. Sciama commented the apparent constancy of $G$:
\bq
`... then, local phenomena are strongly coupled
to the universe as a whole, but owing to the small effect of local
irregularities this coupling is practically constant over the
distances and times available to observation. Because of this
constancy, local phenomena appear to be isolated from the rest of
the universe...'
\eq
Sciama further considered the gravitational potential (eq.~6 there)
\be
\ph = - G \sum_i \frac{m_i}{r_i}= - c^2. \label{Gvalue}
\ee
Despite the inspiring and insightful discussion in the following,
astonishingly he did not consider a spatial variation of $c$, though it seems
a reasonable consequence to relate $c^2$ to the gravitational potential.
As we shall see below, a variable speed of light in combination with (\ref{sci})
 leads to a differential equation that satisfies Dirac's second hypothesis. 
Since Sciama considered the coincidence (\ref{sci}) as approximate, we shall
be able to modify it by a numerical factor.

\subsection{Dicke's `electromagnetic' theory of gravitation}
It was Robert Dicke \cite{Dic:57}\fo{Unfortunately, this paper was published
with the misleading title `Gravitation without a principle of equivalence' which
tells very little about the inspiring content.}
 who first thought of combining 
the dependence (\ref{sci}) with a variable speed of light, apparently 
having been unaware of Sciama's previous efforts. Though Dicke
obviously left this path in favor of the much more prominent
scalar-tensor-theories, we shall investigate this very different
first approach here only. Dicke's proposal belongs to the
`conservative' VSL theories that do not postulate exotic
dependencies of $c$ but widely agree with general relativity (GR)
in the sense that a variable $c$ in a flat background metric
generates a curved space. While
recent VSL theories had to suffer a couple of objections, these do
not apply to the present `bimetric' type,
since the notion of VSL is implicitly present in GR (see \cite{Bro:04}, ref.~70, with numerous
excerpts of GR textbooks). Even before developing the definite version of GR,
Einstein \cite{Einst:11} considered that case.
Dicke realized that the failure of Einstein's 
attempts (see also \cite{Ran:04a}) were due to the neglect of
varying length scales $\lambda$ (Einstein considered varying time
scales only)\fo{It should be noted that though $c$ being a scalar
field here, this theory is not a `scalar' theory {\em coupled to
matter\/} to which Einstein later expressed general caveats; these
reservations were however put into question by \cite{Giu:06}.},
and noted that the classical tests could be described by 
\be
\frac{\delta c}{c} = \frac{\delta  \lambda}{\lambda} + \frac{\delta  f}{f}, 
\ee 
assuming
further $\frac{\delta  \lambda}{\lambda}= \frac{\delta  f}{f}$. Dicke started from
Einstein's idea of light deflection caused by a lower $c$ in the
vicinity of masses \cite{Einst:11}: 
\bq `... that the velocity of
light in the gravitational field is a function of the place, we
may easily infer, by means of Huyghens's  principle, that
light-rays propagated across a gravitational field undergo
deflexion'. 
\eq 
Dicke introduced therefore a variable index of
refraction (\cite{Dic:57}, eq.~5) 
\be 
\e= 1+\frac{2 G M}{r c^2}.  \label{dic} 
\ee 
While the second term on the r.h.s. is related to
the gravitational potential of the sun, Dicke was the first to
raise the speculation on the first term having `its origin in the
remainder of the matter in the universe'. In Appendix~A.1, the
reader will find a brief description how Dicke's  tentative theory may  provide
a formulation of spacetime geometry equivalent to GR and
compatible with the classical tests.\fo{\cite{Yil:58} reports a
private communication that Dicke believed the perihelion of
mercury to come out with a (wrong) factor, a problem which seems
to be settled by the calculations of \cite{Put:99}.} In
Appendix~A.2, it will be oultlined how Newton's law of gravitation
arises from Sciama's hypothesis (\ref{sci}) and can be embedded in
Dicke's model.

\paragraph{The Cosmological redshift in Dicke's proposal} is a cornerstone that
distinguishes drastically from standard cosmology. He described the idea as follows:
\bq
`The cosmological principle was taken to be a fundamental assumption of the theory.
Namely, from any fixed position of a Newtonian frame the universe is assumed to be on the
average uniform. This implies that matter is on the average fixed in position
relative to the Newtonian coordinate frame, for motion would introduce a lack of
uniformity as seen by an observer located where the matter would be moving.
In like manner the scalar field variable $\e$ [polarizability of the vacuum]
and matter density must be position  independent.' (\cite{Dic:57}, p.~374 left).
\eq
Though not stated explicitly, the increase of the horizon $R(t)$ must be governed by
$\dot R(t) = c(t)$, since there is no other possibility for a horizon increase:

\bq
`Although all matter is at rest in this model there is a galactic red shift.
With increasing $\e$ [and decreasing $c$] , the photon emitted in the past has more energy than
its present counterpart. This might be thought to cause a "blue shift". However,
a photon loses energy at twice the rate of loss characteristic of an atom, hence
there is a net shift toward the red. 
(\cite{Dic:57}, p.~374 right).
\eq
The decrease in $c$ is due to new masses dropping
into the horizon. The corresponding decrease of length scales appears as
a net expansion which becomes visible
as cosmological redshift.
Astonishingly,
Dicke did not clearly follow that path and derived a total number of particles
proportional to $\e^{\frac{3}{2}}$, in contrast to Dirac II.\fo{There is a brief
 correspondence on this topic \cite{Dic:61}. To fix
that difference, Dicke introduced another quantity $\frac{\e}{\e_0}$ which later
played an important role in the so-called Brans-Dicke theory.}
It seems that this was due to the quite arbitrary assumption in eq.~(94) that
led to the complicated form of (95) 
which turns out to be in conflict with the differential equation
 that will be derived from (\ref{sci}).
Contrarily, we shall see below that the density that arises from Sciama's
assumption (and Dicke's `rest' of the theory)
matches indeed Dirac's second hypothesis on the number of particles.

\section{Dirac-Sciama-Dicke (DSD) cosmology}

\subsection{Units and Measurement} \label{UaM}

In the following, we assume an absolute, Euclidean space\fo{called `Newtonian' by \cite{Dic:57}.}
and an absolute, undistorted 
 time. The time~$t$ and the distances $r$ expressed in this absolute units
however are  mathematical parameters not directly observable. All
time and distance {\em measurements\/} instead are performed in
relative, dynamical units defined by the actual frequencies $f(t)$
and $\lambda(t)$ of atomic or nuclear transitions. These perceived or
relative quantities measured by  means of $f(t)$ and $\lambda(t)$ shall
be called $t'$ and $r'$. 
In that absolute space, all matter is assumed to
be at rest having a uniform density $\r$ (particles per absolute
volume). In the next subsection we shall consider an evolution of
the horizon $R(t)$ (absolute distance) with the assumption\fo{The
time derivative refers to the absolute time.} $\dot R (t) = c(t)$
starting at $R(t=0)=0$ {\em everywhere\/} in Euclidean space. To
obtain (arbitrarily chosen) time and length scales for the
absolute units, we define $\lambda_0 >0 = \lambda(t_0 >0)$ by the condition
$\frac{4}{3} \pi \r \lambda_0^3 = 1$. Equivalently, we may say the
horizon $R(t_0) = \lambda_0$ contains just one particle. $\dot R (t_0)
= c(t_0) =: c_0$ is then the speed of light at $t=t_0$ in absolute
units and we may define the frequency $f(t_0)=f_0$ by the
identity\fo{To be precise, $t_0$ coincides with $\frac{1}{f(t_0)}$
only by a factor, since we did not introduce further assumptions
on $R(t)$ for the period $0 < t < t_0$. For the choice of units,
this factor does not do any harm. Physically, the (only
reasonable) definition of measurable time by atomic and nuclear
transitions is not possible as long as the transition is not
completed at $t=t_0$.} $ \lambda_0 f_0 = c_0$.

\subsection{Temporal evolution}

Expressing Dicke's index of refraction in (eq.~\ref{dic}) as $\e=\frac{c+\delta c}{c}$
and taking into account the smallness of $\delta c$, with  $\delta c^2= 2 c \delta c$ we may write
\be
\frac{c^2+\delta c^2}{c^2}= 1+\frac{4 G M}{r c^2}. \label{dic2}
\ee
Slightly modifying Sciama's proposal (\ref{sci}) we use $\frac{c^2}{4 G} =\sum_i \frac{m_i}{r_i}$,
leading to\footnote{Sciama explicitly (\cite{Sci:53}, p.~38 below) allowed such a factor.} 
\be
1+\frac{\delta c^2}{c^2}=1+\frac{\frac{M}{r}}{\sum \frac{m_i}{r_i}}.  \label{dicsci}
\ee
Since $\delta$ indicates the difference of values far from and nearby the sun,
we compare the l.h.s. and r.h.s in (\ref{dicsci}) with 
$\frac{M}{r} =\delta \sum \frac{m_i}{r_i}$ and assume all elementary particles 
to have the same mass ($m_i=1$). After integration and cancelling of the arising 
logarithms, this leads to a spatiotemporal dependency of the speed of light
\be
c(\vec r, t)^2= \frac{c_0^2}{\sum_i \frac{1}{|\vec r_i -  \vec r |}}, \label{sci0} 
\ee
 whereby the sum is taken over all particles $i$ and $|\vec r_i -  \vec r |$  
 denoting the  (time-dependent) distance
to particle $i$, measured in absolute units.
I shall abbreviate (\ref{sci0}) as $c_0^2/c^2= \S$ for simplicity\fo{To be in
precise agreement with section~3.1, the sum $\S$ should be replaced
by $\S+1$, since one particle is visible at $t_0$. Since we shall consider only
large values of $\S$ in the following, this will be omitted.}.
The expansion rate $c(t)$
depends therefore on the number of visible particles  and will decrease
while the horizon increases.
Without loss of generality, $r=0$ is assumed, thus we also shall use  the approximation
\be
\S \approx \int_{0}^{R} \frac{ 4 \pi \r r^2 dr}{r} = 2 \pi \r R^2. \label{sum}
\ee
Keeping in mind that $\dot \r =0$, after inserting $\dot R(t) = c(t)$,  (\ref{sci0}) transforms to
\be
\dot R(t)^2= \frac{c_0^2}{2 \pi \r R(t)^2}, 
\ee
which after taking the square root, reduces to  the simple
form
\be
\frac{d}{dt} R(t)^2=const.
\ee
with the solution\fo{This
describes approximately the evolution, since detailed assumptions for $0 < t < t_0$
cannot be given.}
\be
R(t) \sim t^{\frac{1}{2}}; \ \ c(t) \sim t^{-\frac{1}{2}}. \label{dsdevol}
\ee
This evolution is the central difference to FL cosmology 
with $\frac{d}{dt} R(t)= c = const.$\fo{In the matter-dominated epoch.}

\subsection{Change of measuring rods}
Since
the locally observed speed of light $c' = \lambda' f'$ is a constant\fo{$c'=299792458 m/s$ is used
for the SI definition.}, the agreement
with the classical tests of GR (see appendix~A.1)
requires $\lambda \sim t^{-\frac{1}{4}}$ and $f \sim t^{-\frac{1}{4}}$,
that means both wavelengths and frequencies of atomic transitions become smaller during
the evolution of the universe. The  intervals $\tau$ we actually use to measure time change according to
$\tau = t^{\frac{1}{4}}$. Since for the relative, measured time $t'$ the condition
$t' \tau = t \tau_0$ holds ($\tau_0=1$ by definition),
the relative time $t' = \frac{t}{\tau}$ shows a dependence $t' \sim t^{\frac{3}{4}}$ 
(mind that constancy, $\tau \sim t^0$ would lead to the usual $t' \sim t^1$).
The measuring value of the perceived
epoch is $t' = 10^{39}$ now\fo{To avoid fractional exponents, we shall approximate
$10^{40}$ by $10^{39}$.}, therefore the `true' epoch, in absolute units, must be
$t= 10^{52}$ at present.

\paragraph{Dimensionful units and change of further quantities.}

The change of time and length scales has further consequences.
Firstly, all measurements of velocities and accelerations will be affected.
This is already clear for those arising in atoms, otherwise
the scale-defining decline of wavelengths and frequencies could not happen. Thinking in
absolute units, the same particles, undergoing smaller accelerations, have
an apparent {\em inertial mass\/} which accordingly increases.
Developing further this principle of measurement with dynamical scales,
almost all dimensionful physical units turn out to have a time evolution,
thus we may imagine the dependency directly `attached' to a unit like $m$ or $s$. This
eases to find the consistent trend but also elucidates why the change
of physical quantities may be hidden at a first glance in conventional physics.
A list of the respective change of physical quantities for the static case has
already been given by \cite{Dic:57}, p.~366. A corresponding overview is given below in
Table~1. All quantities at $t_0$ are normalized to 1.

\subsection{Observational consequences}

\paragraph{Cosmological redshift.}
As Dicke (\cite{Dic:57}, p.~374) points out, in the context of a
VSL light propagation the following properties hold: $\nabla c$
with approximately $\dot c=0$ affects $\lambda$, while $f$ remains
unchanged. Vice versa, when $\nabla c$ vanishes, $\dot c$ will
change $f $ and leave $\lambda$ constant. Therefore, assuming an
isotropic DSD universe while analyzing the large-scale evolution,
a propagating photon
will change its frequency only, while $\lambda$ is kept fixed. 

\vspace{0.5cm}
\begin{tabular}{|lccr|}
\hline
Quantity    &  Symbol  & evolution $t^{\g}$ & present epoch   \\
\hline
abstract time  & t & $t^{1}$ & $10^{52}$   \\
Horizon  & R & $t^{\frac{1}{2}}$ & $10^{26}$  \\
Speed of light   & c & $t^{-\frac{1}{2}}$  & $10^{-26}$  \\
wavelengths  & $\lambda$ & $t^{-\frac{1}{4}}$  & $10^{-13}$   \\
frequencies   & f & $t^{-\frac{1}{4}}$  & $10^{-13}$    \\
actual time interval  & $\t$ & $t^{\frac{1}{4}}$  & $10^{13}$     \\
velocities   & v & $t^{-\frac{1}{2}}$  & $10^{-26}$  \\
accelerations   & a & $t^{-\frac{3}{4}}$  & $10^{-39}$  \\
perceived Horizon $R^{'}$  & $\frac{R}{\lambda}$ & $t^{\frac{3}{4}}$ & $10^{39}$  \\
perceived epoch t' & $\frac{t}{\tau}$ & $t^{\frac{3}{4}}$ & $10^{39}$  \\
particles  & N & $t^{\frac{3}{2}}$  & $10^{78}$    \\
perceived particle density  & $\r{'}$ & $t^{-\frac{3}{4}}$  & $10^{-39}$    \\
masses & $m$ & $t^{\frac{3}{4}}$  & $10^{39}$    \\
\hline
\end{tabular}\\

\vspace{0.3cm}
Table~1.\\

Consider now a photon emitted at $t_1$ with $c(t_1) = \lambda (t_1) f(t_1)$, in brief
$c_1 = \lambda_1 f_1$. It is detected
later at $t_2$ when other photons ($*$) of the same atomic transition obey
$ \lambda_2^{*} f_2^{*} =c_2^{*} $ with
\be
c_2^{*} = c_1 (\frac{t_2}{t_1})^{-\frac{1}{2}}; \ \
\lambda_2^{*} = \lambda_1 (\frac{t_2}{t_1})^{-\frac{1}{4}}.
\ee
Since the arriving photon still has $\lambda_2 = \lambda_1$ (and $c_2 = c_2^{*}$), it will appear redshifted by the factor
\be
(z+1):= \frac{\lambda_1} {\lambda_2^{*}}= (\frac{t_2}{t_1})^{\frac{1}{4}}.
\ee
Its frequency decreased by $\frac{f_2}{f_1}= (z+1)^{-2}$ with respect to emission,
but is lower only by $\frac{f_2}{f_2^{*}}= (z+1)^{-1}$ with respect to other photons $(*)$ generated at $t_2$.

\paragraph{Dirac's second hypothesis on the total number of particles.}
Since we have assumed an  Euclidean space with constant density $\r$
 in which the horizon increases according to
$R(t) \sim t^{\frac{1}{2}}$ (eq.~\ref{dsdevol}),
for the total number of visible particles
\be
N(t) = \rho V(t) = \frac{4}{3} \pi \r R(t)^3 \sim t^{\frac{3}{2}} \label{2ndH}
\ee
holds. Taking into account that the perceived time shows the dependency
$t'  \sim t^{\frac{3}{4}}$, Dirac's second hypothesis
\be
N(t) \sim t'^{2}
\ee
follows. Of course, the same result is obtained considering
the shortening of length scales $\lambda \sim t^{-\frac{1}{4}}$
causing the perceived horizon to be at the relative distance of
$R'= \frac{R}{\lambda} \sim t^{\frac{3}{4}}$. Then for the number of particles
\be
N= \frac{4}{3} \pi \r'  R'(t)^3 \sim \r' t^{\frac{9}{4}} \sim \r' t'^{3}
\ee
holds, which coincides with (\ref{2ndH}) because $\r' \sim t^{\frac{3}{4}} \sim t'^{-1}$,
an equivalent form of Dirac's second observation.

\paragraph{A possible apparent constancy of the gravitational constant $G$.}

There is some observational evidence  \cite{Uza} against a
temporal variation of $G$.
In the DSD evolution developed above however,
Dirac's postulate of a variation of $G$ turns out to be premature.
First we have to ask what observational evidence supports $\dot G \approx 0$. Exemplarily,
we consider the absence of increasing radii in the Earth-moon\footnote{An anomaly related to this
issue was reported by \cite{Ste:03}.} and the Sun-Mars orbit
(e.g., \cite{WTB:04, Rae:79}), since these are the most
simple ones to discuss.
In the DSD picture, frequencies and wavelengths of atomic transitions
contract according to $f \sim \lambda \sim t^{-\frac{1}{4}}$. Hence, in the classical limit
of orbiting electrons, Bohr's radius\footnote{which is equal to the de-Broglie wavelength
of the orbiting electron divided by $2 \pi$.} has to decline like $r_b \sim \lambda \sim t^{-\frac{1}{4}}$
and the respective centripetal acceleration according to $a_z  \sim t^{-\frac{3}{4}}$.
On the other hand, the gravitational acceleration is proportional to $\nabla c^2$.
Since all gradients are taken with respect to the dynamic units
$\lambda \sim t^{-\frac{1}{4}}$, they appear bigger by the factor $t^{\frac{1}{4}}$, while
$c \sim t^{-\frac{1}{2}}$. Therefore, the gravitational acceleration
 (see appendix~A.2) $a_g \sim t^{-\frac{3}{4}}$
has at least the dependence required for a decrease of the radius $r \sim t^{-\frac{1}{4}}$
in a sun-planet orbit.  This  contraction synchronous with length scales
would result in an apparent absence of any change in distance;
two-body systems, the `planetary clocks', would run slower and contract their orbits in the same manner
as the atomic clocks do.
This result is still in agreement with Kepler's 2nd law, since the angular
momentum $\vec l =m \vec v \times \vec r$, with $m \sim t^{\frac{3}{4}}$,
$v \sim t^{-\frac{1}{2}}$ and $r \sim t^{-\frac{1}{4}}$ yields a time-invariant quantity
even in absolute Euclidean units. From other considerations (see appendix~A.1)
there are good reasons to assume Planck's constant $h$, whose units correspond to $\vec l$,
to be unchanged in time. 

Contrarily to the speed of light, the factor $\frac{F_e}{F_g}$ will yield different measuring values
dependent on the epoch, and therefore the measuring value of $G$, too. The experimental bounds of
absolute $G$ determinations by far do not exclude such a possibility. The commonly expected 
constancy of $G$ and the underlying assumptions of the respective observations 
must be reconsidered in DSD cosmology.

\section{Discussion}

\paragraph{Dirac's hypotheses and agreement with GR phenomenology.}

The most convincing property of DSD cosmology seems the agreement 
with Dirac's large number hypothesises. In particular, also the
 second one is obtained while providing a mechanism for an apparent constancy of $G$,
 which has been used as an argument against Dirac's first hypothesis so far. Mach's principle is 
fully encompassed while the cosmological redshift becomes an intrinsic necessity
in DSD cosmology. 
A critical point to be evaluated further will be the agreement of the underlying
tentative gravity model with GR from a theoretical and experimental point of view. 
For the latter, as far as the classical tests are concerned, DSD cosmology 
does not seem to predict any differences to GR. However, general covariance 
can hardly be achieved since a minute variation of the gravitational constant
is suggested (see A.2). If ever, a consistent formulation must be obtained along the
flat-space formulations of GR, the bimetric theories. 
Though there is a long history (e.g. \cite{Wil:21,Ros:40, Deh:60}), 
the representations in terms of a spatially
varying speed of light (e.g. \cite{Put:99, Arm:04, Bro:04, Bro:05, Bro:05a}) have to 
gain yet broad acceptance.\fo{For possible experimental tests, see \cite{Con:07}.}

In general, there is a wide-ranging observational agreement with conventional cosmology
due to the dynamics of physical units, whose relations to each other change so slowly
that observational differences, if any, remain minute.
This conjecture has still to be verified for the impact of DSD gravity on
electrodynamics,
since nobody would expect $c_{GR}$ to be different from $c_{EM}$ (see \cite{Eli:03} for a
systematic review of the different meanings of $c$). Though Dicke \cite{Dic:57}, p.~372 already
proposed in an explicit way how to modify Maxwell's equations, we cannot go into details here.

Energy conservation is no longer a valuable condition for the evolution of the universe.
Taking a general
perspective, this is not heavily surprising, because energy is a concept introduced
to describe the time-independency of physical laws.\footnote{Conceptual problems
of this kind are addressed in \cite{Bay:06}.}
While this is true for the snapshot
of the universe we are observing, the clumping of matter suggests that the universe is
anything but stationary. Though the differential equation $\frac{d}{dt} R^2= const.$ seems
to be a simple principle, a general formulation, possibly by
means of a Lagrangian, has still to be given.

\paragraph{The flatness and horizon problem.}

These cosmological puzzles triggered the rivival of modern VSL 
theories (\cite{Alb:99, Bar:99}, for an overview \cite{mag:03}) which 
provided  solutions alternative to inflation.
Here we restrict to the fact that
flatness is closely related to the observation of the approximate
coincidence (\ref{mach}). As it is evident from (\ref{sci}),
the apparent $G$ must have an according value in the same order of
magnitude. It does not make sense at the moment to relate DSD
predictions to the WMAP data which set tight bounds on flatness
like $\Om = 1 \pm 0.02.$ Given that deviations from $\Om = 1$ at
primordial times should cause huge deviations at present,
approximate coincidence following from Sciama's ansatz (\ref{sci})
is a step towards an explanation of `flatness'.

In Friedman-Lemaitre cosmology, gravity acts as a contracting
force which slows down the Hubble expansion. It is precisely that
slowdown  that causes new masses to drop into the horizon and
raises the question how masses, without having causal contact,
could show a highly uniform behavior like the CMB emission.
Contrarily, in DSD cosmology, since all matter is initially at
rest, masses attract due to gravitational interaction, but this
does not affect the apparent redshift. Consequently, the problem
of slowing down the  `expansion' does not even arise.

\paragraph{Cosmic Microwave Background.}
It is interesting to investigate the impact of the present proposal for
the WMAP data of the  cosmic microwave background. According to common cosmology,
the CMB is a signal from the recombination period at $z \approx 1100$, commonly
assumed to be 380000 years after the big bang. Assuming an nonuniform
evolving time like in DSD cosmology, $\frac{\lambda'}{\lambda}-1 = z  \approx 1100$ corresponds, since
$\lambda \sim t^{-\frac{1}{4}} $, to an epoch of $t=3 \times \ 10^{42}$, while
at present $t=10^{53}$ holds. Measured in units of
the  `local' time, that epoch corresponds to $t' =10^{31}$, i.e. about one year.
This is a dramatic difference and must carefully be compared to the observations.
A calculation of the power spectrum that has to take into account different
temperature and density assumptions, would be premature at this stage.
As far as the amplitude of CMB fluctuations is concerned, one expects much tinier
fluctuations in DSD cosmology since there is much more time left for the fluctuations to
evolve to galaxies. One should keep in mind that before the COBE data had analyzed,
much greater fluctuations were expected, a riddle which was resolved in the following
by assuming corresponding dark matter fluctuations.

\paragraph{Big Bang.}
Though we were not able to discuss the details shortly after $t=0$, some substantial
differences to FL cosmology should be noted. The absolute scale  $\lambda_0$ was defined above
by the condition of a single particle being contained in the horizon. If one assumes
this particle to be a baryon, its rest energy corresponds to the 
zero energy $E_0$ of a particle closed in a quantum well
of the size of the horizon:
$E_0 = \frac{h}{t_0} = \frac{h c}{\lambda_0} \approx m_p$. In this case, the evolutionary 
equation in absolute units writes as\fo{Given that $m_p$ increased by a factor
$10^{40}$ in the meantime (see table~1),
the perceived $\frac{h}{m_b}$ accordingly decreased to the actual value
$5 \cdot 10^{-7} \frac{m^2}{s}$.} $\frac{d}{dt} R^2 = \frac{h}{m_p}$. In general, a density equal to the
density of nuclear matter seems to require much less extrapolation of physical laws
than the densities that arise in FL cosmology shortly after the big bang.

\section{Outlook}
The present proposal based on the ideas of Dirac, Sciama and Dicke is a first  
framework for a cosmology based on a tentative alternative gravity model.
 Regarding the quantity of observations in agreement with a theoretical
framework, the DSD proposal is unable to compete with
standard FL cosmology with its currently accepted $\Lambda$CDM model. DSD cosmology may only
gain importance if one is disposed to raise doubts to (1) the validity of the standard model
with its considerable extrapolation of the laws of nature and an increasing number of free parameters
(2) the suggestion of the standard model that Mach's principle and Dirac's enigmatic hypotheses being
just numerical coincidences (3) the conviction of the constants of nature being fixed
but arbitrary numbers; this last condition seems the most entrenched one.
However the idea that we are observers living inside a prison of dynamic measuring instruments,
which in first approximation cause a blindness for the perception of change, is certainly
not unfamiliar. 

\paragraph{Acknowledgement.}
The author thanks for various inspiring discussions with Karl Fabian.

\begin{appendix}

\section{Appendix}
\subsection{Tentative VSL formulation of GR}
It is a known feature of GR that in a gravitational field clocks run slower and a shortening of
measuring rods occurs with respect to clocks and rods outside the field.
Defining $c$ with respect to the latter scales, one can equivalently say that
$c$ is lowered\fo{This is sometimes called `non-proper' speed of light, see
\cite{Ran:04a}.} in the gravitational field (cfr. \cite{Wil}, p.~111, and 
\cite{Bro:04}, ref.~70). 
Based on that point of view, an equivalent descriptions of GR by means of 
VSL theories can be tried, e.g. \cite{Put:99} (and references above) which is known as `polarisable
vacuum' (PV) representation  of GR .
The only necessary postulate is that since $c= f \lambda $ and
$\delta c = f \delta \lambda + \lambda \delta f$, the relative change 
\be 
\frac{\delta c}{c}=\frac{\delta f}{f}+\frac{\delta
\lambda}{\lambda}, \label{cfl} 
\ee 
is equally\fo{In 1911, Einstein
considered fixed $\lambda$'s only, thus $\frac{\delta c}{c}=\frac{\delta f}{f}$,
being in accordance with the `Newtonian' value that failed to
match the famous data of Eddington's eclipse observation in 1919.}
 distributed to $f$ and $\lambda$, that means $\frac{\delta f}{f}=\frac{\delta \lambda}{\lambda}$.
We shall denote the quantities outside the gravitational field as
$c, \lambda, f$ and the lower quantities in the field as $c^*, \lambda^*, f^*$.
Hence, in a gravitational field, clocks run slower by the relative
amount 
\be 
f^* = \a^{-1} f; \ \a:= (1+\frac{G M}{r c^2}) 
\ee 
and
wavelengths $\lambda$ shorten by the same factor $\a$: $\lambda^* = \a^{-1}
\lambda$, which is a well-known result of GR. According to (\ref{cfl}),
$c^* = \a^{-2} c$ has to be lowered by 
$\a^2 \approx (1+\frac{2G M}{r c^2})$ in a weak-field approximation. While that idea has
first been developed by \cite{Dic:57}, the results of \cite{Put:99}, sec.~III,
suggest that all classical tests of GR can
 be described in this manner, see also \cite{Unz:05}, sec.~3-5\fo{I do not uphold any longer
the considerations in sec.~6.}.
To give an example, we briefly describe the gravitational redshift
of the sun \cite{Sni:72}.

One can imagine the process as follows: consider a
photon travelling from the gravitational field of the sun to the
earth (with approximately zero gravity). Starting at $f^*, \lambda^*, c^*$, while travelling
 it keeps its (lowered) frequency $f^*$. At earth, i.e. outside
the gravitational field where
$c = \a^2 c^*$ is higher by the double amount (\ref{cfl}), the photon
has to adjust its $\lambda$, and raise it with respect to the value $\lambda^*$ at
departure. Since the adjustment $\a^2$ to $c$ overcompensates the originally
lower $\lambda^* = \a^{-1} \lambda$, we detect the photon as gravitationally 
redshifted with $ \a^2 \lambda^* = \a \lambda$.

\paragraph{Change of measuring rods.}
Time and length measurements naturally affect accelerations
(${a \sim \a^{-3}}$), and surprisingly, masses, too. Photon and rest masses, $h f$ and
$m c^2$ have to behave in the same manner, and since $f \sim \a^{-1}$,
$c^2 \sim \a^{-4}$,  $m \sim \a^3$ must hold. This is in agreement
with Newton's second law according to which masses have to be proportional to inverse
accelerations. An overview on the relative change of various quantities inside the gravitational field
(cfr. \cite{Dic:57}, p.~366 and \cite{Deh:60}) is given in Table~2 below. 
$\a$ denotes a factor of $(1+\frac{G M}{r c^2})$:

\vspace{0.5cm}

\begin{tabular}{|lcccr|}
\hline
Quantity & symbol &unit & rel. change &\\
\hline
speed of light & c  & $\frac{m}{s}$ & $\a^{-2}$ &\\
Frequency  & f  & $\frac{1}{s}$ & $\a^{-1}$ &\\
Time  & t  & $s$ & $\a$ &\\
Length  & $\lambda$  & $m$ & $\a^{-1}$ &\\
Velocity  & v  & $\frac{m}{s}$ & $a^{-2}$ &\\
Acceleration  & a  & $\frac{m}{s^2}$ & $\a^{-3}$ &\\
Mass  & m  & $kg$ & $\a^3$ &\\
Force  & F  & $N$ & $\a^0$ &\\
Pot. energy  & $E_p$  & $Nm$ & $\a^{-1}$ &\\
Ang. mom.  & l  & $\frac{kg m^2}{s}$ & $\a^0$ &\\
\hline
\end{tabular}\\
\vspace{0.3cm}

Table~2: Relative change of quantities inside the gravitational field.

\subsection{Newton's law from a variable $c$.}

Once time and length measurement effects of GR are described
by a spatial variation of $c$,  all gravitational phenomena should
be encompassed by the same framework. However, $c \sim \a^{-2}$
requires (eqns.~\ref{dic}, \ref{dic2}) in first approximation
\be 
\delta (c^2)= 2 c \delta c = -\frac{4 G M}{r}. \label{4pot} 
\ee 
This leads to a Newtonian
gravitational potential of the form 
\be \ph_{Newton} = \frac{1}{4}c^2, 
\ee 
which differs\footnote{An early investigation on the Mach-Sciama approach 
(\cite{Tre:72}, p.~93) deduces $\frac{1}{3}c^2$ , which seems to be incompatible with the
Newtonian limit.} by a factor 4 from Sciama's potential.
Sciama's proposal was however always considered as approximate by
the author (\cite{Sci:53}, p.~38 below). Since (\ref{sci0}) 
\be
c(\vec r)^2= \frac{c_0^2}{\sum_i \frac{1}{|\vec r_i -  \vec r |}},
\label{gpot} 
\ee for the acceleration of a test mass 
\be \vec a (\vec r)
= - \frac{1}{4} \nabla c(\vec r)^2 = \frac{c_0^2}{4 \S^2}
\sum_i \frac{\vec r_i -  \vec r}{|\vec r_i -  \vec r |^3} 
\ee
holds. Assuming without loss of generality $|\vec r| = r = 0$ and substituting $c_0^2$,
\be
a= \frac{c^2}{4 \S} \sum_i \frac{\vec e_i}{r_i^2} \label{newton}
\ee
follows, yielding the inverse-square law. Thus $c_0$ does not appear any more
and the Newtonian force is perceived in the local, dynamic units. As one easily verifies,
(\ref{newton}) does not depend on the units in which  masses and distances are measured.
Thus (\ref{gpot}), while setting $r=0$,  may be rewritten in SI quantities 
(with an new reference $c_{n}$):
\be
c^2= \frac{c_{n}^2}{\sum_i \frac{m_i}{r_i}}, \label{gpot2}
\ee
The `gravitational constant' is then given by the quantity
\be
G= \frac{c^2}{4 \sum_ i \frac{m_i}{r_i}},  \label{GG2}
\ee
in accordance with \cite{Sci:53}. From (\ref{GG2}) and the assumption of an homogeneous universe,
elementary integration over a spherical volume yields
$\sum \frac{m_i}{r_i} \approx \frac{3 m_u}{2 r_u}$, and therefore
\be
m_u \approx \frac{c^2 r_u}{6 G}
\ee
holds, which is in approximate agreement with the amount of baryonic matter.
 
\end{appendix}

\end{document}